\documentclass{MICE}
\usepackage{lineno}
\usepackage{bibentry}
\nobibliography*
\usepackage{times}
\usepackage{bm}         % bold maths
\usepackage{amsmath,amssymb,bm,slashed} % need for subequations
\usepackage{amssymb}    % extra math symbols
\usepackage{graphicx}   % need for figures
\usepackage{verbatim}   % useful for program listings
\usepackage{color}      % use if color is used in text
\usepackage{subfigure}  % use for side-by-side figures
\usepackage{hyperref}   % use for hypertext links, including those to external documents and URLs
\usepackage{enumerate}

\usepackage{fmtcount}   % Allow counters to be printed with padding

\usepackage[square,comma,numbers,sort&compress]{natbib}

\modulolinenumbers[5]
\begin{document}

\thispagestyle{empty}

\begin{tabular}{p{0.175\textwidth} p{0.5\textwidth} p{0.225\textwidth}}
  \hspace{-0.8cm}\leftline{RAL-P-2015-009}                                 &
  \centering{ Muon Ionization Cooling Experiment}                  &
  \rightline{\today} 
\end{tabular}
\vspace{-1.0cm}\\
\rule{\textwidth}{0.43pt}

\renewcommand{\thefootnote}{\ifcase\value{footnote}\or$\dagger$\or*\or
$\ddagger$\or\#\or$\dagger\dagger$\or**\or$\ddagger\ddagger$\or \#\#\or $\infty$\fi}
\begin{center}
  {\bf
    {\LARGE Pion contamination in the MICE muon beam} \\
  }
  \vspace{0.2cm}
  The MICE collaboration\footnote{Authors are listed at the end of this paper.} \\
  \vspace{-0.0cm}
\end{center}
\renewcommand{\thefootnote}{\arabic{footnote}}
\setcounter{footnote}{0}

\makeatletter

% Quantum mechanical notation
\newcommand{\bra}[1]{\ensuremath{\langle #1 |}}   % Bra vector
\newcommand{\ket}[1]{\ensuremath{| #1 \rangle}}   % Ket vector
\newcommand{\bigbra}[1]{\ensuremath{\big\langle #1 \big|}}   % Bra vector
\newcommand{\bigket}[1]{\ensuremath{\big| #1 \big\rangle}}   % Ket vector
\newcommand{\amp}[3]{\ensuremath{\left\langle #1 \,\left|\, #2%
                     \,\right|\, #3 \right\rangle}}  % QM amplitude
\newcommand{\sprod}[2]{\ensuremath{\left\langle #1 |%
                     #2 \right\rangle}}  % QM scalar product
\newcommand{\ev}[1]{\ensuremath{\left\langle #1 %
                     \right\rangle}} % Expectation value
\newcommand{\ds}[1]{\ensuremath{\! \frac{d^3#1}{(2\pi)^3 %
                     \sqrt{2 E_\vec{#1}}} \,}} % Spatial integral
\newcommand{\dst}[1]{\ensuremath{\! %
                     \frac{d^4#1}{(2\pi)^4} \,}} % Space-time integral
\newcommand{\tr}{\text{tr}}
\newcommand{\sgn}{\text{sgn}}
\newcommand{\diag}{\text{diag}}
\newcommand{\BR}{\text{BR}}

% Miscellaneous commands
\renewcommand{\vec}[1]{{\mathbf{#1}}}
\renewcommand{\Re}{{\text{Re}}}
\renewcommand{\Im}{{\text{Im}}}
\newcommand{\iso}[2]{{\ensuremath{{}^{#2}}\ensuremath{\rm #1}}}
\newcommand{\eps}{{\ensuremath{\epsilon}}}
\newcommand{\draftnote}[1]{{\bf\color{red} \MakeUppercase{#1}}}
\newcommand{\panm}[1]{{\color{blue} #1}}
\providecommand{\abs}[1]{\lvert#1\rvert}
\providecommand{\norm}[1]{\lVert#1\rVert}

\def\parenbar{\mathpalette\p@renb@r}
\def\p@renb@r#1#2{\vbox{%
  \ifx#1\scriptscriptstyle \dimen@.7em\dimen@ii.2em\else
  \ifx#1\scriptstyle \dimen@.8em\dimen@ii.25em\else
  \dimen@1em\dimen@ii.4em\fi\fi \offinterlineskip
  \ialign{\hfill##\hfill\cr
    \vbox{\hrule width\dimen@ii}\cr
    \noalign{\vskip-.3ex}%
    \hbox to\dimen@{$\mathchar300\hfil\mathchar301$}\cr
    \noalign{\vskip-.3ex}%
    $#1#2$\cr}}}

%
%.. Sanjib's commands:
\providecommand{\anmne}{\mbox{$\bar\nu_{\mu} \rightarrow \bar\nu_e$}} 
\providecommand{\nmne}{\mbox{$\nu_{\mu}\rightarrow\nu_e$}} 
\providecommand{\anm}{\mbox{$\bar\nu_\mu$}} 
\providecommand{\nm}{\mbox{$\nu_\mu$}}
\providecommand{\nue}{\mbox{$\nu_e$}} 
\providecommand{\ane}{\mbox{$\bar\nu_e$}} 
\providecommand{\enu}{\mbox{$E_\nu$}}
\providecommand{\piz}{\mbox{$\pi^0 $}}
\providecommand{\pip}{\mbox{$\pi^+$}} 
\providecommand{\pim}{\mbox{$\pi^-$}}

\parindent 10pt
\pagestyle{plain}
\pagenumbering{arabic}                   
\setcounter{page}{1}

\begin{quotation}

\noindent
The international Muon Ionization Cooling Experiment (MICE) will perform a systematic investigation of ionization cooling with muon beams of momentum between 
140 and 240\,MeV/c at the Rutherford Appleton Laboratory ISIS facility. The measurement of ionization cooling in MICE relies on the selection of a pure 
sample of muons that traverse the experiment. To make this selection, the MICE Muon Beam is designed to deliver a beam of muons with less than $\sim$1\% 
contamination. To make the final muon selection, MICE employs a particle-identification (PID) system upstream and downstream of the cooling cell. 
The PID system includes time-of-flight hodoscopes, threshold-Cherenkov counters and calorimetry. The upper limit for the pion contamination measured in this paper is $f_\pi < 1.4\%$ at 90\% C.L., including systematic uncertainties.  Therefore, the MICE Muon Beam is able to meet the stringent pion-contamination requirements of the study of  ionization cooling. 
\end{quotation}

\section{Introduction}
\label{Sect:Intro}

The international Muon Ionization Cooling Experiment (MICE)~\cite{Blondel:2003},
at the ISIS facility of the Rutherford Appleton Laboratory (RAL), will
demonstrate the principle of ionization cooling as a technique for
reducing the phase-space volume occupied by a muon beam.
Ionization-cooling channels are required for neutrino factories
\cite{Koshkarev:1974, Geer:1998, Ozaki:2001bb,Alsharoa:2002wu,Blondel:2004ae,Choubey:2011zz} 
and muon colliders \cite{Gallardo:1996aa,Tikhonin:2008pw,Geer:2010zz, Ankenbrandt:1999as}, since this is the only known technique that
can achieve the required cooling performance within the short muon lifetime.

Ionization cooling \cite{Neuffer:1983,Neuffer:1983jr} 
is accomplished by passing the muon beam
through a low-$Z$ material (the ``absorber''), in which it loses
energy via ionization, reducing both the longitudinal and
transverse components of momentum.
The lost energy is restored by accelerating the beam such that the
longitudinal component of momentum is increased, while the transverse
components remain unchanged.
The net effect is to reduce the emittance of the beam.
Beam transport through the absorbers and accelerating structures is
achieved using a solenoid-focusing lattice.
Cooling factors of between 2 and 50 are required for recent neutrino factory designs
\cite{Choubey:2011zz,Delahaye:2013jla}, but much greater ($\sim$10$^6$) six dimensional (6D) cooling is required for a muon collider.

Three lithium hydride (LiH) absorbers, two radio-frequency (RF) cavities and two Focus Coil
solenoid magnets %(figure \ref{fig:MICElayout}) 
will be used to reduce the transverse emittance of the muon beam by up to 8\%, depending on the beam configuration \cite{Hanlet:2014kga}. 
The goal of MICE is to measure the 
transverse normalised emittance before and after the cooling cell with an accuracy of 0.1\%. This is achieved using two spectrometers consisting of scintillating-fibre trackers inside solenoid magnets \cite{Ellis:2010bb}. 
Any unidentified contamination in the muon beam from pions and electrons can affect the accuracy of the measurement of the muon-beam emittance. Electrons are identified 
using a time-of-flight (TOF) system \cite{Bertoni:2010by} and an Electron--Muon Range (EMR) detector \cite{Lietti:2009zz,Adams:2015eva} after the cooling channel. Pions in the beam are also identified
by the TOF system, two aerogel Cherenkov detectors \cite{Cremaldi:2009zj}, a preshower calorimeter (Kloe-Light or KL) \cite{Bogomilov:2012sr} and the EMR. In order to achieve 0.1\% accuracy in the emittance measurement, it is essential that the muon sample selected in the beam has a pion contamination below $\sim$1\%. 
The particle identification should achieve a pion rejection factor between 10 and 100, so a pion contamination in the beam of  $\sim$1\% should reduce the misidentified pion contamination in the muon sample to less than 0.1\%, required to achieve the physics goals.
The pion contamination of the MICE Muon Beam was measured in dedicated data-taking runs in order to qualify the muon beam and to ensure
that MICE can achieve its stated physics goals \cite{Bogomilov:2012sr,Adams:2013lba}. 

The paper is organised as follows: a brief description of the MICE  
experiment is included in Section \ref{Sect:MICE_Apparatus}, the MICE Muon Beam is described briefly in Section \ref{Sect:Beam}, the analysis method is described in 
Section \ref{Sect:Contamination}  and the results and systematic errors are given in Section \ref{Sect:Results}, 
followed by a brief conclusion (Section \ref{Sect:Conclusions}).

\section{MICE apparatus}
\label{Sect:MICE_Apparatus}

\begin{figure}[htb]
\begin{center}
\includegraphics[width=\linewidth]{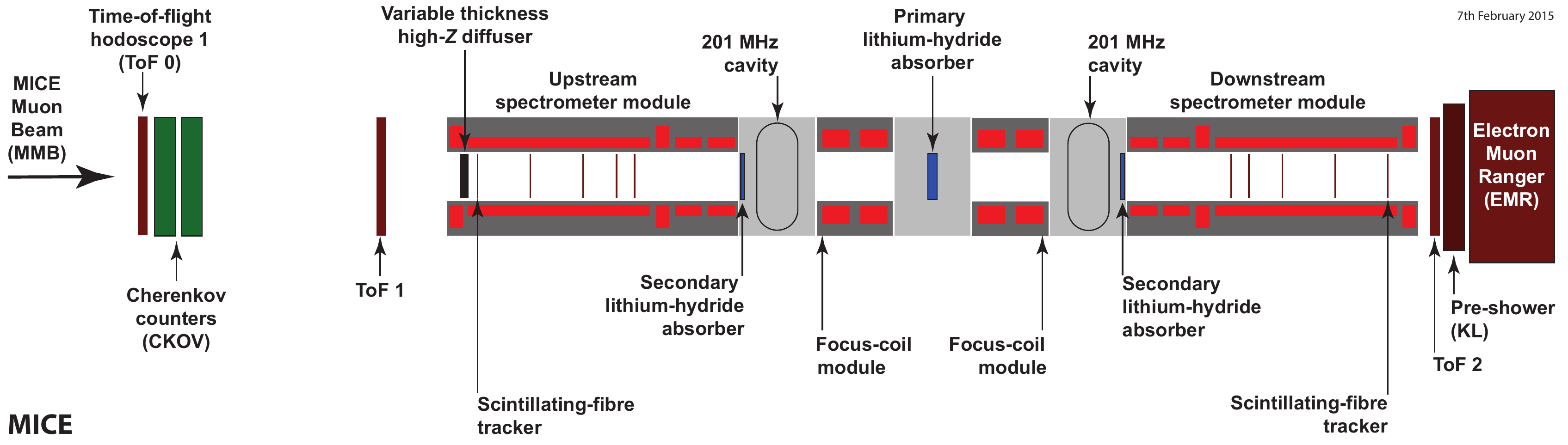}
 \end{center}
%\vskip -3cm 
\caption{
Schematic view of the MICE experiment, with three LiH absorbers (one primary absorber in the centre and two secondary absorbers), two RF cavities and
two focus-coil magnets that define the MICE optics, sandwiched between two identical trackers, inside superconducting  solenoids.}
 \label{fig:MICElayout}
\end{figure}

The MICE experiment, shown schematically in figure~\ref{fig:MICElayout}, is similar to the
cooling channel for the International Design Study for the Neutrino Factory \cite{Choubey:2011zz}, and differs from the
original cooling channel design in \cite{Ozaki:2001bb}. It consists of one primary lithium-hydride (LiH) absorber, two secondary absorbers, two focus coils and
two 201\,MHz RF cavities that provide an accelerating gradient of  $\sim$10.3\,MV/m. The two superconducting focus-coil modules ensure that the transverse betatron function is minimised at the position of the absorbers, thereby increasing the cooling performance of the channel.

For a muon beam entering the cell with a nominal momentum of 200\,MeV/$c$ and 4D normalised  emittance $\epsilon_N = 5.8 \pi $\,mm$\,\cdot$\,rad,
a 6\% cooling effect is expected \cite{MICE-Note-452}. Conventional emittance-measurement techniques based on beam-profile monitors cannot achieve the required precision, so MICE has been designed as a single-particle experiment, in which each muon is measured using state-of-the-art particle detectors and the bunched muon-beam  is reconstructed offline \cite{Adams:2013lba}. The tracking spectrometers~\cite{Ellis:2010bb} upstream and downstream of the cooling cell consist of
scintillating-fibre tracking modules inside solenoid magnetic fields that measure the emittance before and after the cooling cell. These are required to 
measure the normalised transverse emittance, $\epsilon_N$, with a precision $\sigma_{\epsilon_N} / \epsilon_N \sim 0.1\%$.

The MICE instrumentation includes a PID system that allows a pure muon
beam to be selected. The PID system consists of scintillator time-of-flight $x/y$ hodoscopes TOF0, TOF1 and TOF2~\cite{Bertoni:2010by}  
read at both ends of each scintillator slab by fast Hamamatsu R4998 photomultiplier (PMT) tubes \cite{Bonesini:2011}, and two threshold Cherenkov counters 
Ckova and Ckovb \cite{Cremaldi:2009zj}. The  TOF system is required to tag electrons and pions in the muon beam with a rejection factor
exceeding 99\%. Furthermore, the precision of the TOF time-measurement must be sufficient to allow the phase at which the muon enters the RF 
cavities to be determined to 5$^\circ$. To satisfy these requirements, the resolution of each TOF station must be $\sim$50\,ps. 
The TOF resolutions obtained are 55~ps for TOF0, 53~ps for TOF1 and  50~ps for TOF2  \cite{datax,Bonesini:2012}.

The two Cherenkov detectors have been designed to guarantee muon-identification purities better than $99.7$\% in the momentum range 
210\,MeV/$c$ to 365\,MeV/$c$ \cite{Sanders:2009vn}. The TOF and the Cherenkov systems work in combination with the 
upstream tracking spectrometer~\cite{Ellis:2010bb} to identify the particles \cite{Bogomilov:2012sr,Bonesini:2013ila}. 

TOF2 \cite{tof2} and a calorimeter system allow muon decays to be identified and rejected downstream of the cooling cell. 
The calorimeter system for MICE consists of the
KLOE--Light (KL) lead-scintillator sampling calorimeter, similar to the KLOE design \cite{Ambrosino:2009zza} but with thinner lead foils,
designed to serve as a preshower for the EMR totally-active scintillating detector. 
The main roles of the KL and EMR detectors are to distinguish muons from decay electrons and pions. 
In this paper, however, the pion contamination of the MICE Muon Beam is measured on a statistical basis using 
data taken before the MICE tracking spectrometers and the EMR
were installed. The analysis is accomplished by combining the TOF velocity information with the KL calorimetric information.  The KL calorimeter is composed of scintillating fibres and extruded lead foils, with an active volume of  93 $\times$  93  $\times$ 4 cm$^3$. It has 21 cells, and the light from its scintillating fibres is collected by 42 Hamamatsu R1355 PMTs. 
The PMT signals are sent via a shaper module to 14 bit CAEN V1724 flash ADCs.  The shapers stretch the signal in time in order to match the flash ADC sampling rate. 
A detailed description of KL is given in \cite{Bogomilov:2012sr}.

\section{MICE Muon Beam}
\label{Sect:Beam}
The required normalised transverse emittance  range of the MICE Muon Beam is $3 \leq \epsilon_N \leq 10 \pi $\,mm$\,\cdot$\,rad, with mean momentum in the range
$140 \leq p_\mu \leq 240$\,MeV/$c$ and a root-mean-squared (RMS) momentum spread of $\sim$20\,MeV/$c$. A pneumatically operated ``diffuser", 
consisting of tungsten and brass irises of various thicknesses, is placed at the entrance to the upstream spectrometer solenoid in order to generate the
 required range of emittance. In order to perform the muon-emittance measurement with the required accuracy of 0.1\%
it is essential to limit the pion and electron contamination of the muon sample to less than 0.1\%. This is achieved by designing a muon
beam with $\sim$1\% contamination and then by using the PID system to further identify electrons and pions passing through.

\begin{figure}[htb]
%\vskip -2cm 
 \centering
  \includegraphics*[width=\linewidth]{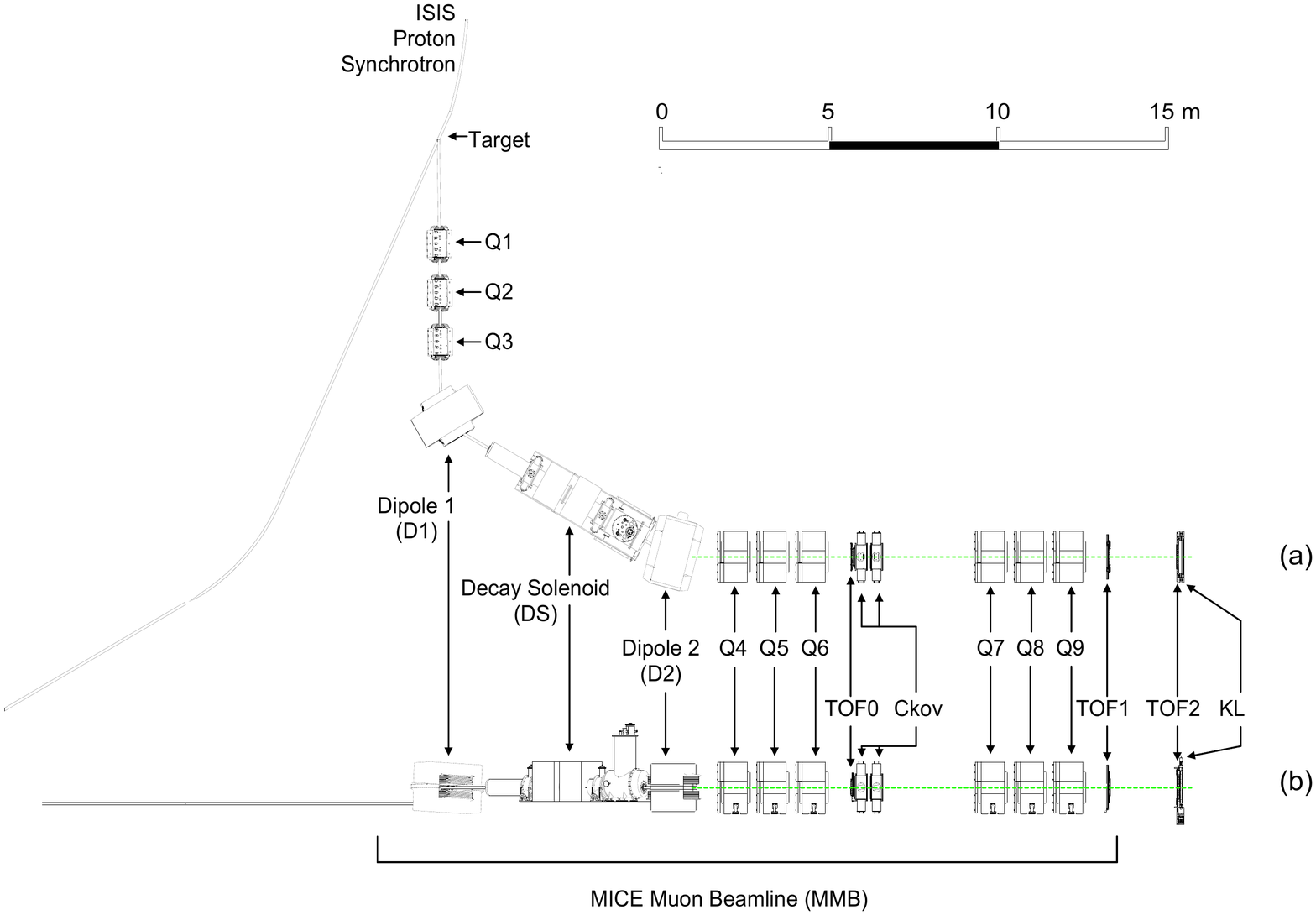}
  %\vspace*{-0.5cm}
  \caption{(a) Top view of the MICE Muon Beam and its instrumentation for the pion contamination measurement. (b) Side view of the MICE Muon Beam.}
 \label{fig:Beamline}
\end{figure}

The design of the MICE Muon Beam is briefly summarised here (see figure \ref{fig:Beamline}) and is reported in detail in  \cite{Bogomilov:2012sr}.
Pions produced by the momentary insertion of a titanium target \cite{target:2013} into the 800~MeV ISIS proton beam are captured using a 
quadrupole triplet (Q1--3) and transported to a  first dipole magnet (D1), which selects
particles of a desired momentum bite into the 5~T decay solenoid (DS).  Muons produced by pions decaying in the DS are momentum-selected using a
second dipole magnet (D2) and focussed onto the diffuser by a quadrupole channel (Q4--6 and Q7--9). By capturing pions of transverse 
momentum up to $\sim$70\,MeV/$c$,  and increasing their path length by deflecting them onto helical trajectories, the decay solenoid increases
the probability of muon capture between D1 and D2 by an order of magnitude compared to a simple quadrupole channel.
In positive-beam running, a borated polyethylene  absorber of variable thickness is inserted into the beam just downstream of the DS in order to suppress the
high rate of protons that are produced at the target \cite{Blot:2011zz}.

The composition and momentum spectra of the beams delivered to MICE are determined by the interplay between the two bending magnets D1 and
D2. In normal (``$\pi \to \mu$ mode,'' or ``muon'') operation, D2 is set to half the momentum of D1, selecting backward-going muons in the pion rest
frame and producing an almost pure muon beam. Pions of high momentum that do not decay may be present in the beam and it is this small contamination that is the focus of the measurement presented in this paper.  In the absence of a precise momentum measurement from the spectrometer, single-particle pion identification is not possible, 
since the particle mass cannot be obtained by combining the momentum with the velocity obtained from either the TOF or Cherenkov detectors.  
Therefore, the measurement has been performed on a statistical basis using the KL and  TOF information.
Alternatively, by setting $p_{D1} \simeq p_{D2}$, a mixed beam containing pions, muons and electrons is obtained. This  ``calibration mode'' is used to calibrate 
the particle identification detectors and is used in the analysis to provide ``templates" for the particle-identification performance of the KL and TOF detectors to be determined.

The nominal values of the beam momenta, $p_{\mu}$, are those evaluated at the centre 
of the central LiH absorber, taking into account the energy lost by the particles along the muon beam in 
the TOF and Cherenkov detectors, the proton absorber (for positive polarity beams), the diffuser and the air along the particle trajectories.
For example, a momentum at D2, $p_{D2} = 238$\,MeV/$c$, implies a momentum value $p_{\mu} = 200$\,MeV/$c$ at the centre of the central absorber. 

Data were taken in December 2011 with the muon beam shown in figure \ref{fig:Beamline},  including the upstream 
TOF0 and TOF1 detectors, Cherenkov detectors and the downstream TOF2 and KL detectors, which were operated in a temporary 
position about 2~m downstream of TOF1. The precise distances between TOF0 (TOF1) and TOF1 (TOF2) in this configuration
are respectively 773.3 cm and 198.8 cm. The correspondence between beam momentum at various points in the MICE beam for the 
muon-beam configuration and the different calibration beams used in this analysis is summarised in table \ref{Table:calibruns}.

\begin{table}
\centering
\caption{Summary of runs used in this analysis. The muon runs correspond
to a nominal setting $(\varepsilon_N, p_\mu)= (6 \pi$\,mm$\,\cdot$\,rad, 200 Mev/$c$). Reported momenta are
at the entrance of the quoted detectors.  \label{Table:calibruns}}
%\vskip 0.2cm
\begin{tabular}{|c|c|c|c|c|}
\hline
\multicolumn{5}{|c|}{Muon runs} \\ \hline
$p_{D2}~({\rm MeV}/c)$&  $p_{TOF0}~({\rm MeV}/c)$ & $p_{TOF1}~({\rm MeV}/c)$ &
$p_{TOF2}~({\rm MeV}/c)$ &\# events ($10^3$) 
\\ \hline
238 & 220 & 204 & 190      & 270 \\
\hline
\multicolumn{5}{|c|}{Calibration runs} \\ \hline
$p_{D2}~({\rm MeV}/c)$&  $p_{TOF0}~({\rm MeV}/c)$ & $p_{TOF1}~({\rm MeV}/c)$ &
$p_{TOF2}~({\rm MeV}/c)$ &\# events ($10^3$) 
\\ \hline
222 &   217 & 194 & 181 &     195 \\
258 &   254 & 231 & 219 &      235 \\
280 &   276 & 254 & 242 &      167 \\
294 &   290 & 268 & 257 &      354 \\
320 &   316 & 295 & 284 &       265 \\
362 &   358 & 337 & 326 &       448 \\ 
\hline
\end{tabular}
\end{table}

\section{Method for determining the contamination in the MICE Muon Beam}
\label{Sect:Contamination}

\begin{figure}[htb]
  \begin{center}
\includegraphics[width=0.49\linewidth]{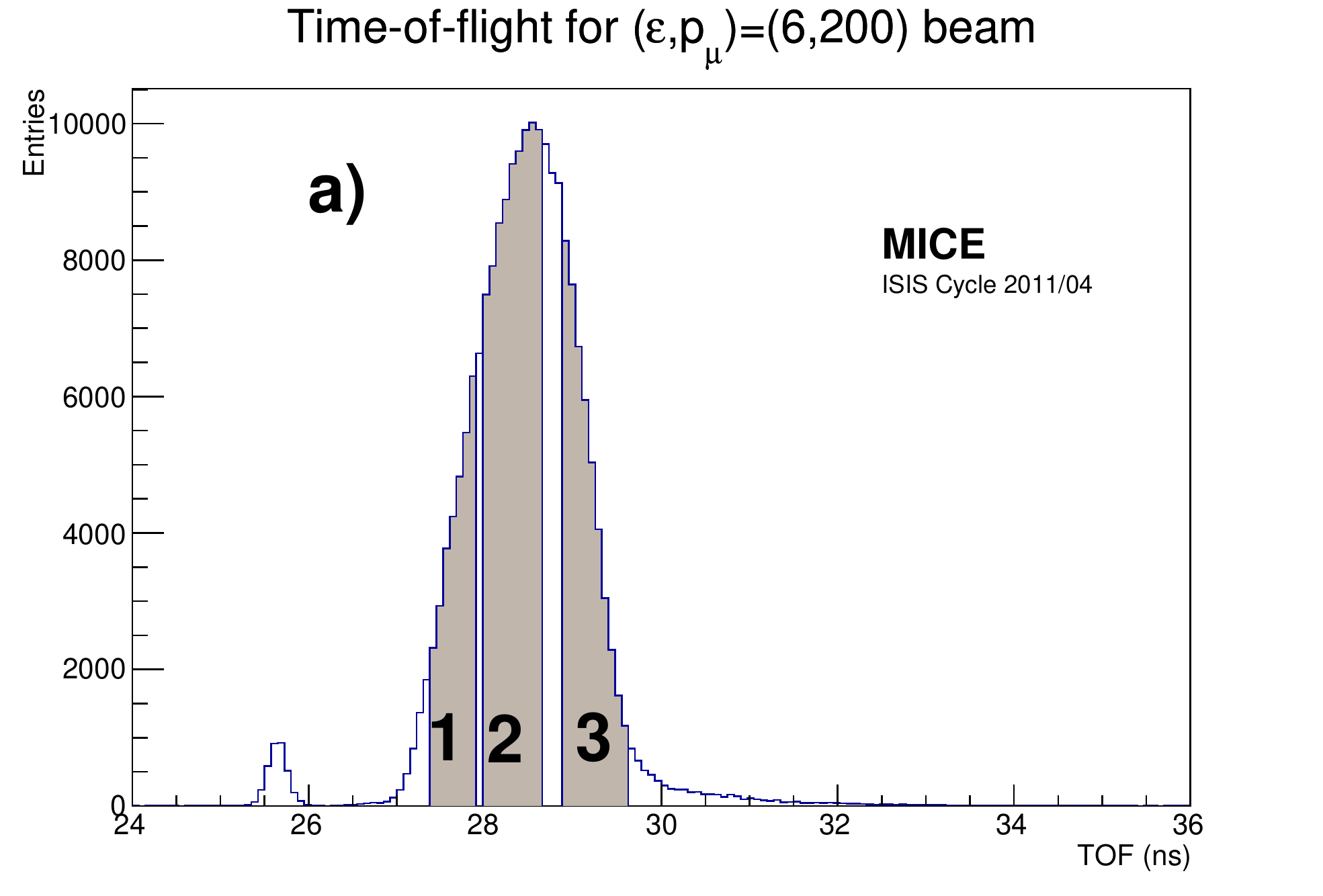}
    \includegraphics[width=0.49\linewidth] {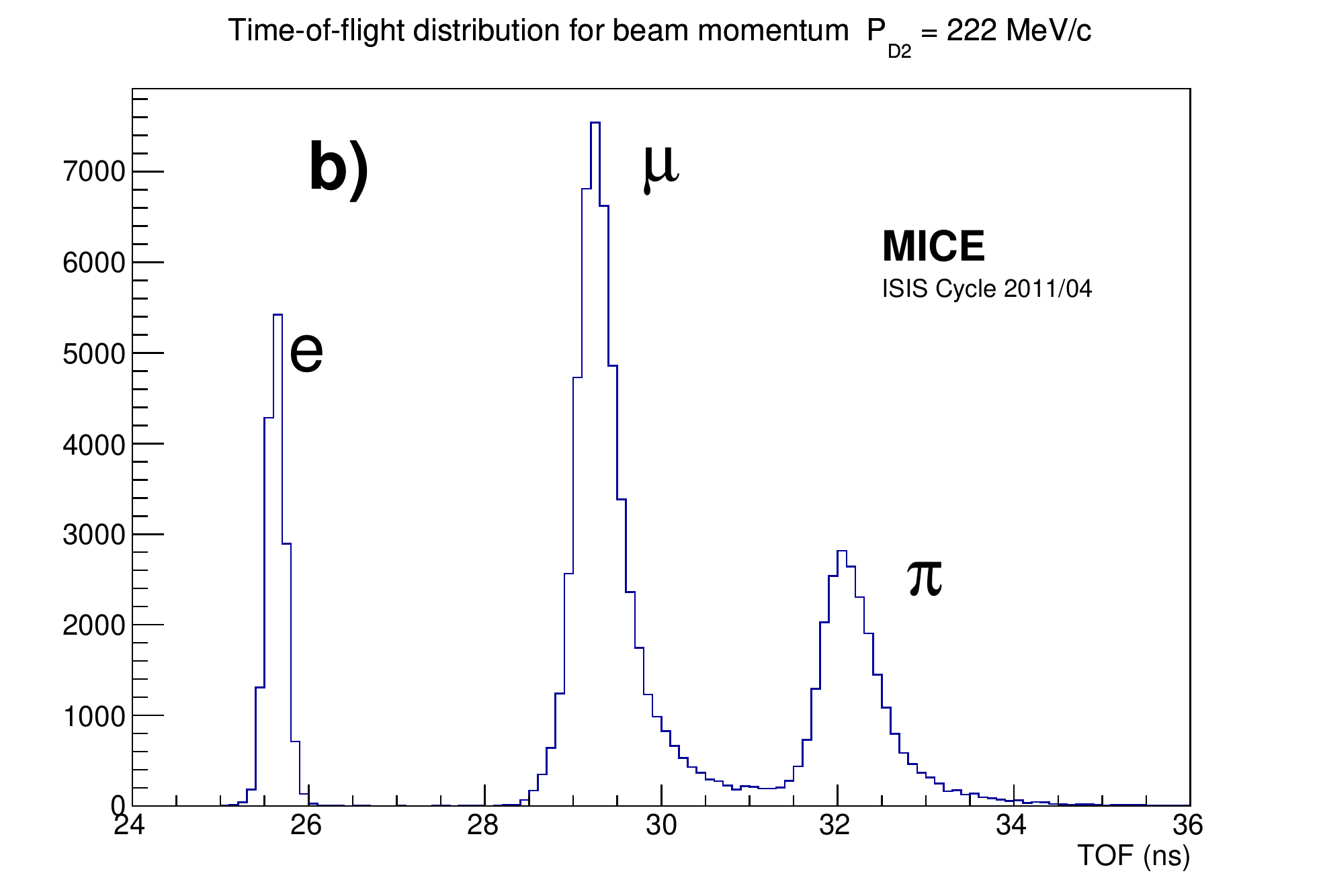}
  \end{center}
  \caption{
    (a) Time of flight distributions between TOF0 and TOF1 for a positive muon  beam  with a nominal momentum of 200\,MeV/$c$ (the left peak is due to electrons). The labels 1, 2 and 3 in the muon peak refer to the three time-of-flight intervals, highlighted in grey, used in the analysis. 
    (b) Positive ``calibration'' beam taken with $p_{D2} = $222\,MeV/$c$, showing clear electron, muon and pion peaks.
  }
  \label{tof}
\end{figure}

The purpose of the analysis presented here is to determine the pion contamination of the MICE Muon Beam by using information from the TOF system and the KL detector. Figure \ref{tof} shows distributions of the time-of-flight of particles between
TOF0 and TOF1, with a positive $\pi \rightarrow \mu$ beam of nominal momentum 200\,MeV/$c$ (figure \ref{tof}a) and with a  calibration beam of $p_{D2} \simeq$ 222\,MeV/$c$
(figure \ref{tof}b). An electron peak is observed that is well separated from the main muon peak,  but the level of the pion contamination under the muon peak
cannot be determined from this distribution alone, as the muon and pion distributions  overlap. However, for the 222\,MeV/$c$ calibration beam, the electron, muon and pion peaks are well separated by 
their time-of-flight. The muon peak in the $\pi \rightarrow \mu$ beam is broader than that of the calibration beam, since the muons selected by D2 originate from pion decays in a range of angles in the backward hemisphere of the pion rest frame \cite{Bogomilov:2012sr}.

\begin{figure}[htb]
%\vskip -3cm
\begin{center}
\includegraphics[width=.49\linewidth]{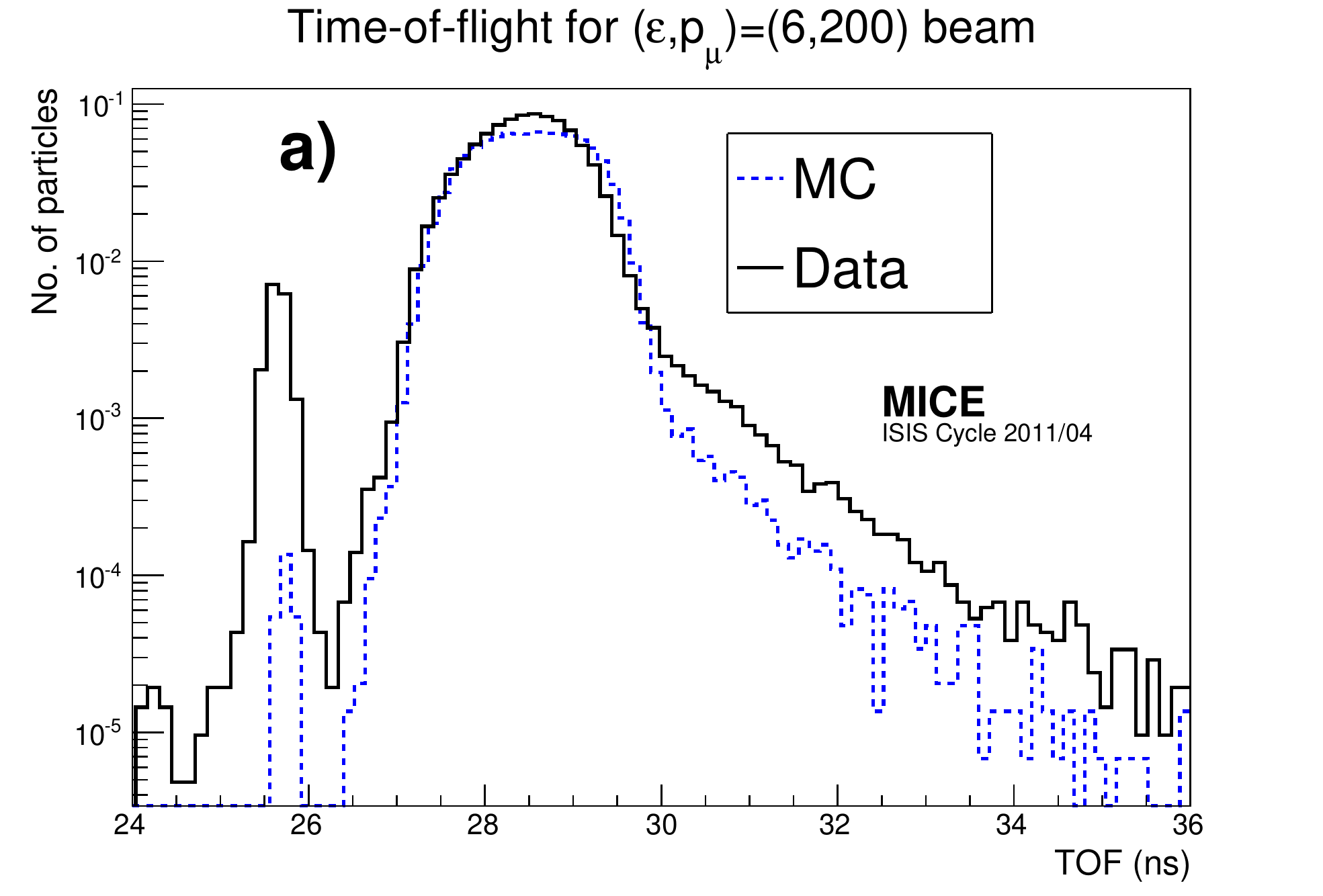}
\includegraphics[width=.49\linewidth]{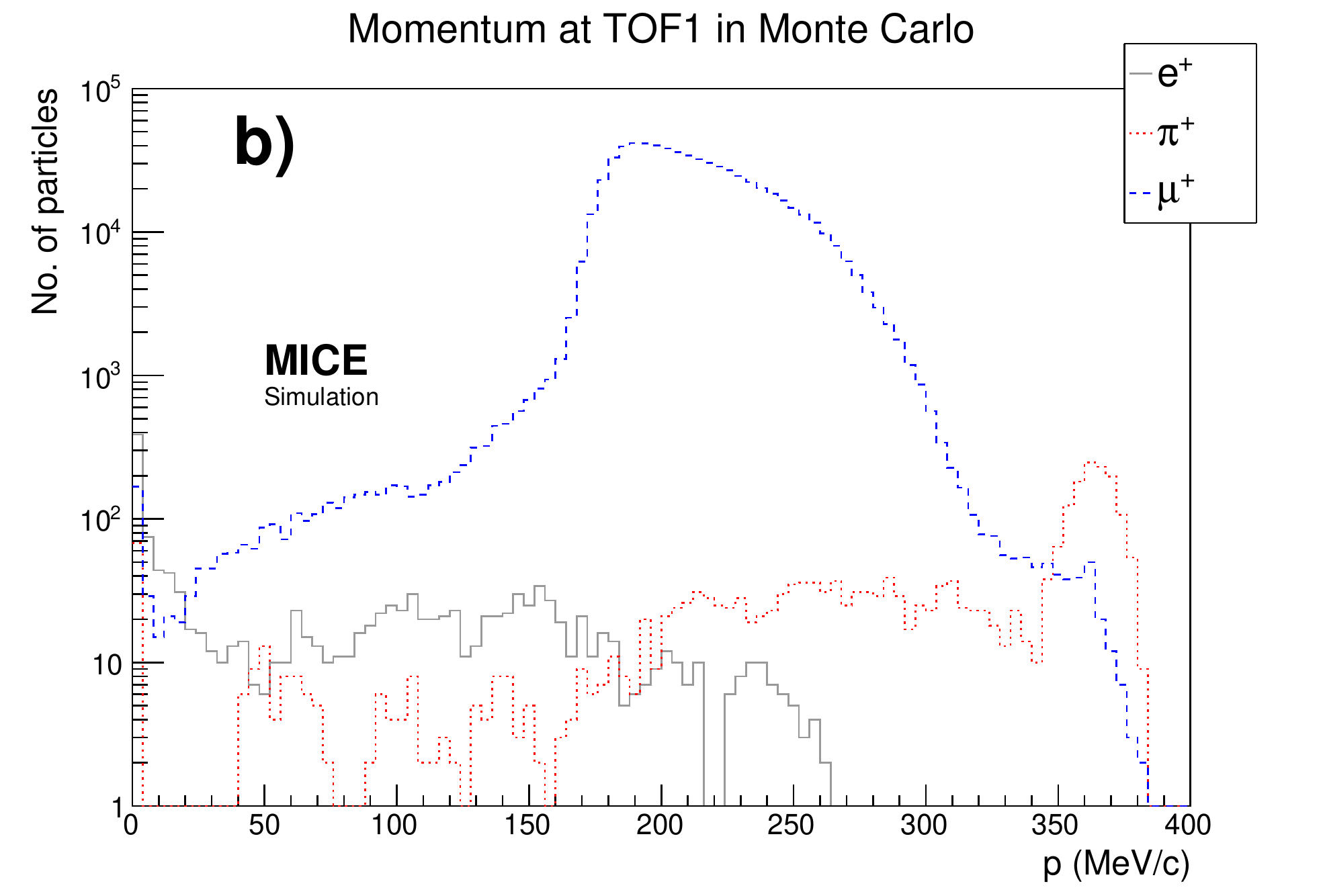}
\end{center}
\caption{(a) Time-of-flight distributions between TOF0 and TOF1 for data and  Monte Carlo simulation for a 
$ 6 \pi$\,mm$\,\cdot$\,rad positive muon 
beam with nominal beam momentum $p_{\mu}=200$\,MeV/$c$. (b) Momentum distribution 
for beam particles at TOF1 for 
a simulated positive $6 \pi$\,mm$\,\cdot$\,rad beam at 200\,MeV/$c$ (the time-of-flight between TOF0 and TOF1 is required to satisfy $26.2 <$ TOF $< 32$~ns).}
\label{fig:sim}
\end{figure}

The pion contamination under the muon peak was estimated using the G4beamline simulation package \cite{G4beamline} and the MICE Applications User Software (MAUS) 
package \cite{MAUS} to simulate detector response. Figure \ref{fig:sim}a compares distributions of flight time from TOF0 to TOF1
for reconstructed positive-beam data and corresponding Monte Carlo simulations of $ 6 \pi$\,mm$\,\cdot$\,rad positive muon beams with nominal beam momentum 
$p_{\mu}=200$\,MeV/$c$. The electron contamination is underestimated in the Monte Carlo simulation because the simulation does not transport particles that interact in the material at the edge of the beam acceptance, but charge exchange interactions can produce neutral pions, and these can decay to electrons and positrons in the beam line. Furthermore, the tail of the time-of-flight distribution is also underestimated in the Monte Carlo simulation. Due to these differences between data and Monte Carlo simulation, this pion contamination analysis is purely based on data, and the Monte Carlo simulation is only used to validate the method.

Figure~\ref{fig:sim}b shows the momentum distribution at TOF1 of the electron, pion and muon peaks for the same 
Monte Carlo simulation, showing that the pion contamination under the muon peak is predominantly due to high momentum pions (with a smaller low momentum component) that are selected by the D2 dipole magnet
and are subsequently transported by the beam. 
Since the muon sample and the higher-momentum pions that contaminate it have similar times of flight, the TOF detectors cannot be used to distinguish them from each other.
Therefore, the residual pion contamination in the beam, after the application of time-of-flight requirements suitable for the selection of muons, can only be measured using the spectrum of energy deposited in KL. The pion contamination is a function of the position at which it is measured. According to the G4beamline simulation, the
contamination under the muon peak at TOF0  is estimated to be 1.78\%, reducing to 0.38\% at TOF1 and  0.22\% at KL.

The pion contamination is studied in positive-muon-beam runs with nominal  beam momentum 200\,MeV/$c$ ($p_{D2}=238$\,MeV/$c$)  and with a sample corresponding to approximately $270 \times 10^3$ triggers. The study is performed as a function of the time-of-flight of the beam particles in three distinct time-of-flight intervals 
(referred to below as ``Points 1, 2 and 3'') the choice of which is dictated by the availability of calibration data for which the specified interval is populated mainly by muons or mainly by pions.
Pairs of calibration runs for which muons and pions present time-of-flight values within the same range (see  table \ref{Table:pairs}) are defined for each point and are used to benchmark the KL response  to muons or to pions of given time-of-flight. In figure \ref{tof}a,  the three points are highlighted in grey in the time-of-flight distribution of particles in the MICE Muon Beam.

\begin{table}
\begin{center}
\caption{Paired beam settings for three time-of-flight intervals (``Points").}
\label{Table:pairs}
{\small
\begin{tabular}{|c|c|c|c|}
\hline
 & TOF interval, ns & muons from runs with & pions from runs with\\ 
 &   & P$_{D2}$ (MeV/$c$) & P$_{D2}$ (MeV/$c$)\\ 
\hline
\hline
Point 1    & 27.4 -- 27.9    & 294  &  362\\
\hline
Point 2    & 28.0 -- 28.6    & 258  &  320\\
\hline
Point 3    & 28.9 -- 29.6    & 222  &  280\\
\hline
\end{tabular}}
\end{center}
\end{table} 

\begin{figure}[htb]
\begin{center}
\includegraphics[scale=0.6]{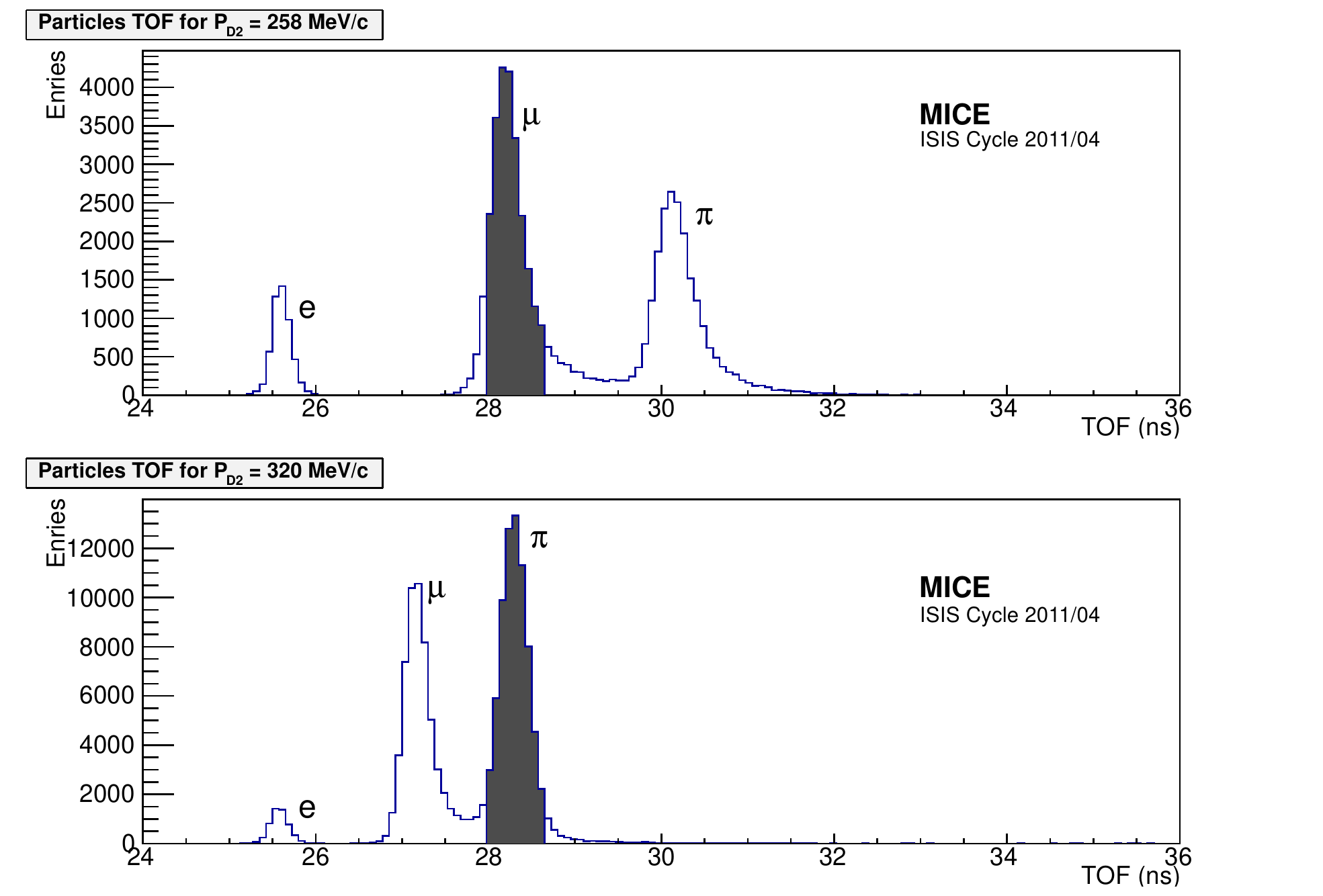}
\caption{Time-of-flight distributions in two paired beam settings. 
The interval 28.0--28.6 ns (shaded) is populated by muons (pions) in the upper (lower) plot.}
\label{Fig:TOFpairing}
\end{center}
\end{figure}

The widths of the intervals were determined by taking into account the overlap regions between the calibration runs.  In each of these time-of-flight intervals 
the spectra of the KL response can be extracted for muons and pions separately from the calibration runs. These spectra are then used as templates for the response to 
muons and pions in that time-of-flight interval for the muon runs.  As an example, figure \ref{Fig:TOFpairing} shows the time-of-flight distributions in two paired beam settings.
The interval 28.0--28.6 ns in the TOF0--TOF1 time-of-flight (point 2) is populated mainly by muons for one beam setting and by pions for the other.

The minimum ionizing responses of muons and pions in the KL are similar, 
but pions can also undergo hadronic interactions, which are visible as a tail in the KL response  to pions. 
The KL response to a particle is defined in terms of the product of the digitised signals from the left and right sides of each scintillator slab divided by their sum: 
$$ADC_{\rm product} = 2\, \frac{ADC_{\rm left}\times ADC_{\rm right}}{ADC_{\rm left}+ ADC_{\rm right}},$$
where the factor of 2 is present for normalisation\footnote{The normalised ADC product is used to compensate for light attenuation in the scintillator and 
to diminish the dependence of the PMT signals on the particle-hit position, since the optical fibres are characterised by two attenuation lengths 
\cite{DiDomenico}.}.

\begin{figure}[htb]
\begin{center}
\includegraphics[scale=0.5]{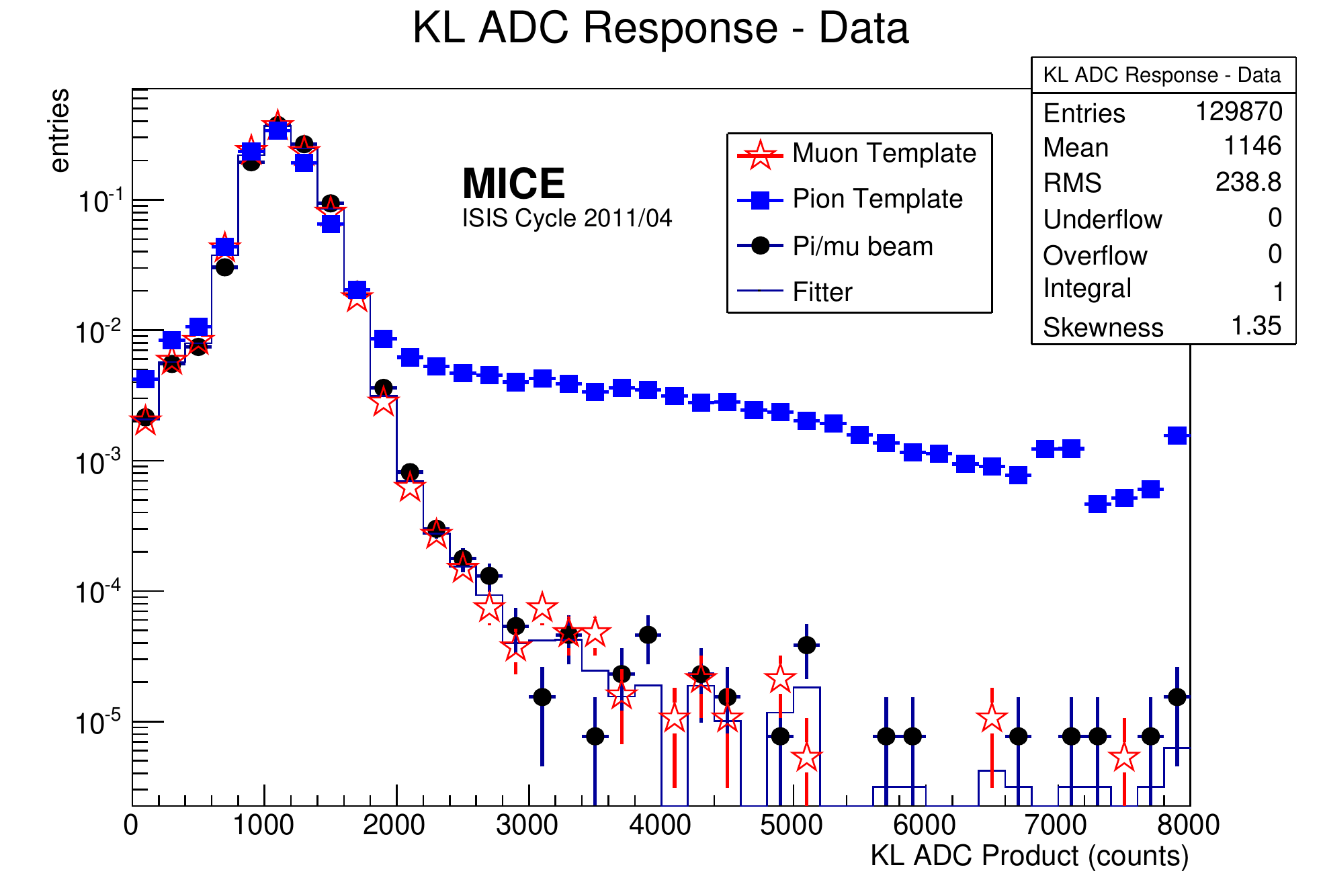}
\caption{Muon template (red stars) and pion template (blue squares) data for the sum of the three TOF data intervals from calibration runs, 
compared to MICE $\pi \rightarrow \mu$ beam data 
(black dots). The histogram is the result of a fit of the $\pi \rightarrow \mu$ beam to the fraction of pions and muons based on the two templates. Plots are normalised to unity.}
\label{Fig:KL_response_data}
\end{center}
\end{figure}

The normalised ADC products are summed for all scintillator slabs in the KL that have a signal above a threshold. 
The KL response to muons and pions in calibration runs and to a particle  mix in the $\pi \rightarrow \mu$ beam mode are added together for the three TOF
intervals (Points 1, 2 and 3) and shown in figure \ref{Fig:KL_response_data}. An additional constraint was imposed that only one track was present in both
the time-of-flight detectors, associated to only one hit in the KL detector.
The distribution for the pions displays a larger tail than that for the muons, due to the presence of hadronic interactions. This feature is used in the following analysis to estimate the contamination on a statistical basis. 

\begin{figure}[htb]
\begin{center}
\includegraphics[scale=0.5]{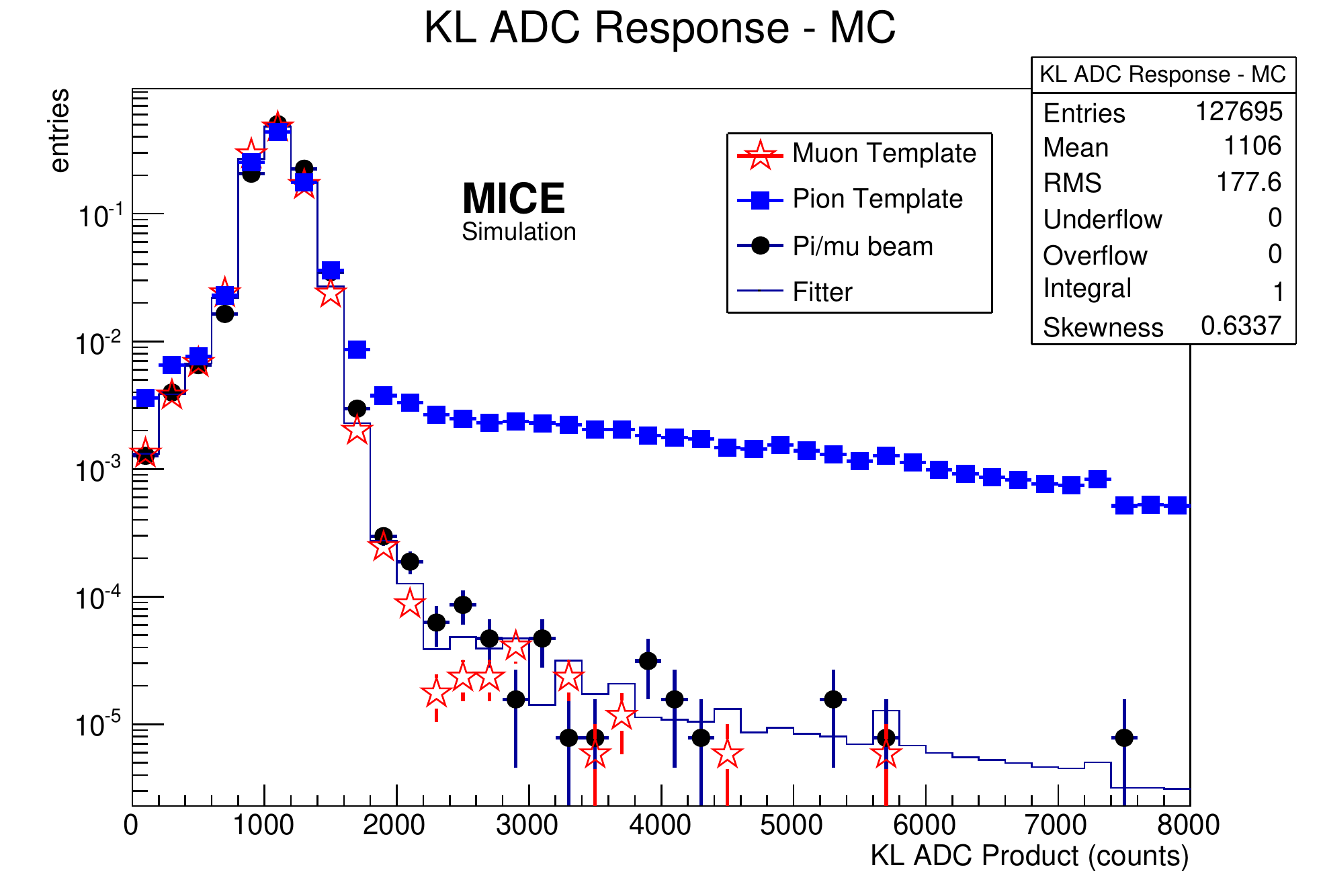}
\caption{Monte Carlo simulation of the muon template (red stars) and pion template (blue squares)  for the sum of the three TOF data intervals, 
compared to the simulated MICE $\pi \rightarrow \mu$ beam
data (black dots). The histogram is the result of a fit of the simulated $\pi \rightarrow \mu$ beam to the fraction of pions and muons based on the two templates. Plots are normalised to unity.}
\label{Fig:KL_response_MC}
\end{center}
\end{figure}

The MAUS simulation of the KL response was fine-tuned in order to match features observed in the data. 
The following features were taken into account:
\begin{itemize}
\item Poisson smearing of the photon count produced in the scintillating fibres and the photoelectrons produced at the photocathode of the PMT;
\item The distribution of photomultiplier gain, assumed to be Gaussian with mean $\sim$2$\times 10^6$ and standard deviation equal to half the gain \cite{PMT_Gain}; and
\item The conversion factors from photoelectrons to ADC counts (250,000\,PE/ADC), from MeV to photoelectrons (0.000125\,MeV/PE), the two-component scintillating-fibre 
attenuation lengths  (2400\,mm and 200\,mm), the scintillating-fibre collection efficiency (3.6\%), the light-guide collection efficiency (85\%) and the photomultiplier-tube quantum efficiency (26\%), in order to obtain $\sim$1060 ADC counts for a minimum-ionizing peak.
\end{itemize}
The Monte Carlo simulation of the KL response to muons and pions for the calibration runs and for the simulated $\pi \rightarrow \mu$ beam are shown in figure \ref{Fig:KL_response_MC}. The features of the simulated Monte Carlo KL response  to pions and muons follow closely that from the data in figure \ref{Fig:KL_response_data}.

The fraction of pions and muons in the $\pi \rightarrow \mu$ beam is extracted by exploiting the information contained in the full KL response spectrum for the sums of the three time-of-flight intervals. The method employs the ROOT TFractionFitter \cite{Bib:ROOT,Bib:fitter} to fit the normalised muon and pion templates to the actual KL spectrum in the MICE data. This was carried out for both the extracted MICE data and for the simulated Monte Carlo distributions for the $6\pi$\,mm$\,\cdot$\,rad, 200~MeV/c $\pi \rightarrow \mu$ beam. The fits for the weighted sum of the three time-of-flight windows (27.4  ns -- 27.9 ns, 28.0 ns -- 28.6 ns, 28.9 ns -- 29.6  ns) are shown as histograms for the data in figure \ref{Fig:KL_response_data}
and for the Monte Carlo simulation in figure \ref{Fig:KL_response_MC}.  The fits take into account both data and template statistical uncertainties
through a standard likelihood-fit method.

\section{Results of the pion contamination in the muon beam and systematic errors}
\label{Sect:Results}
The data from the 6$\pi$\,mm$\,\cdot$\,rad, 200\,MeV/$c$ muon beam encompassing the three time-of-flight windows includes $N_b = 129870$ beam events.  
The fractions of muon and pion events were allowed to converge without any restrictions. The total fitted number of muon events was $N_\mu=130173$, 
which yields $N_\pi = -303\pm509$ pion events, compatible with zero.  Similarly, for the Monte Carlo simulation, the fitted number of muon events 
$N^{MC}_\mu=127772$ was also compatible with the number in the beam $N^{MC}_b = 127695$, which also yielded a number of pions compatible with zero, 
$N_\pi = -77\pm505$.

The Feldman--Cousins likelihood-ratio ordering-procedure \cite{Feldman:1997qc} is a unified frequentist method to construct single- and double-sided confidence intervals for parameters of a given model adapted to data. It provides a natural transition between single-sided confidence intervals, used to define upper or lower limits, and double-sided ones. It is particularly useful near the boundaries of physical regions, while providing a true confidence interval. The Feldman--Cousins procedure was used to extract an upper limit of the pion contamination in the $\pi \rightarrow \mu$ beam at the KL detector position $f_\pi < 0.69\%$ at 90\% C.L. An upper limit for the 
Monte Carlo simulation at the KL position $f^{MC}_\pi < 0.86\%$ at 90\% C.L. was also derived, to be compared to the ``true" pion contamination from the Monte Carlo simulation of 
$0.22\pm 0.01\%$.

The sources of systematic errors considered in this analysis were:
\begin{itemize}
\item  Finer subdivision of the time-of-flight windows;
\item  Shift in the calibration of the time-of-flight windows;
\item  Binning of the KL ADC histograms;
\item  Effects of muon contamination in the pion templates (pion contamination in the muon template was found to be negligible); and
\item  Loosening the constraint that there is only one hit in the KL detector ($N_{KL}=1$) to having one or more hits in KL ($N_{KL}>0$).
\end{itemize}

The systematic errors for both data and the Monte Carlo simulation on the pion contamination are given in table \ref{Tab:FitSys}. The systematic error due to the dependence on the time-of-flight distribution was determined by further subdividing the time-of-flight ranges associated with each point.
Doubling the number of time-of-flight bins varies the fitted pion contamination by 0.18\%. The dependence of the pion-fraction obtained on the time-of-flight calibration
is determined by shifting independently the time-of-flight values in the calibration runs by an amount compatible with the electron peak position ($\pm0.1$~ns). 
This results in a small variation in the pion contamination of 0.04\% for data and 0.28\% for Monte Carlo. The dependence on the histogram binning in
the KL ADC distribution was also assessed by doubling and halving the bin-size to yield a variation in the fitted pion contamination of 0.14\% in data and 0.16\% in simulation.
There is a small bias in the determination of the pion contamination due to the expected muon contamination in the pion template. For example, the nominal value is 25.1\% muons in the pion template for point 1, 26.1\% muons  for point 2 and 26.2\% muons for point 3. Setting the muon contamination in the pion template to zero in the 
Monte Carlo results in a shift in the pion contamination in the $\pi \rightarrow \mu$ beam by 0.03\%. Loosening the number of KL hits from $N_{KL}=1$ 
to $N_{KL}>0$ results in a change in the fit of 0.25\%.

The quadratic sum of the total systematic errors is shown in the bottom row of table \ref{Tab:FitSys}. The total systematic error for the pion contamination is found to be 
0.34\% in data and 0.45\% in Monte Carlo.   These systematic errors are used to obtain the following yields:  $N_\pi = -303\pm509$ (stat)$\pm442$ (syst) for the data and $N_\pi = -77\pm505$ (stat)$\pm 575$ (syst) for the Monte Carlo. The statistical and systematic errors are added in quadrature and the Feldman--Cousins procedure is repeated to extract new upper limits of the pion contamination in the $\pi \rightarrow \mu$ beam at the KL position of $f_\pi < 1.37\%$ at 90\% C.L. including systematic errors.  An upper limit for the Monte Carlo simulation with systematic errors was also derived: $f^{MC}_\pi < 2.06\%$ at 90\% C.L. An analysis using only the TOF and Cherenkov detectors has obtained a comparable limit \cite{MICE-Note-473}.

\begin{table}[hbt]
\centering
\caption{Sources of systematic errors in the evaluation of the pion contamination. }
\label{Tab:FitSys}
\begin{tabular}{|l|l|p{1.5cm}|p{1.5cm}|}
%\toprule
%\midrule
\hline
Effect                       &  Assessment method                  &  \multicolumn{2}{|p{3cm}|}{Absolute Impact on $\pi$ contamination} \\
\hline
%\hline
%\multicolumn{2}{|r|}{}                                             &  Data   & MC  \\
                             &                                     &  Data   & MC  \\\cline{3-4}
Time-of-flight distribution  &  Finer subdivision                  &  0.18\%   & 0.18\%\\
Time-of-flight calibration   &  Shift calibrations by $\pm$0.1 ns  &  0.04\%   & 0.28\%\\
%Fitted range                 &  vary exclusion region              &  0.14\%  &\\
Histogram binning            &  Double/halve bin sizes             &  0.14\%   & 0.16\%\\
Bias due to contamination in templates & Create pure templates in MC & 0.03\%  & 0.03\% \\
Bias in selection            &  Cut KL cell hits $> 0$               &  0.25\%   & 0.25\% \\ 
\hline 
\multicolumn{2}{|r|}{Total}                                         & 0.34\%    & 0.45\% \\ 
%\bottomrule
\hline
\end{tabular}
\end{table}

\section{Conclusions}
\label{Sect:Conclusions}

An upper limit to the pion contamination in the MICE Muon Beam at the position of the KL detector has been determined using precision time-of-flight counters in combination with the KL calorimeter. The measurements were carried out in a variety of time-of-flight windows and the analysis yielded a pion contamination compatible with zero. 
The Monte Carlo expectation for the pion contamination of a $\pi \rightarrow \mu$ beam of 6$\pi$\,mm$\,\cdot$\,rad emittance and 200\,MeV/$c$ nominal momentum is $(0.22\pm0.01)\%$ at the KL.
The upper limit for the pion contamination at the KL position was found to be $f_\pi < 1.4\%$ at 90\% C.L., including systematic errors. This upper limit on the pion contamination in the MICE Muon Beam, combined with the performance of the PID system, meets the experimental requirement.
%\newpage
% ----------------------------------------------------------------
\section*{Acknowledgements}
% ----------------------------------------------------------------

The work described here was made possible by grants from Department of Energy and National Science Foundation (USA), the Instituto Nazionale di Fisica Nucleare (Italy), the Science and Technology Facilities Council (UK), the European Community under the European Commission Framework Programme 7 (AIDA project, grant agreement no. 262025, TIARA project, grant agreement no. 261905, and EuCARD), the Japan Society for the Promotion of Science and the Swiss National Science Foundation, in the framework of the SCOPES programme. We gratefully acknowledge all sources of support.

We are grateful to the staff of ISIS for the reliable operation of ISIS. We acknowledge the use of Grid computing resources deployed and operated by GridPP in the UK, http://www.gridpp.ac.uk/.

\clearpage
\bibliographystyle{utphys}
\bibliography{Concatenated-bibliography}

%\cleardoublepage
\appendix
%========  Author list:
%\vspace{-8mm}
\clearpage
\thispagestyle{plain}
\setlength\parindent{0em}%\noindent
%\section*{The MICE collaboration}

{\large \bf The MICE collaboration} \\

\renewcommand{\thefootnote}{\alph{footnote}}
\setcounter{footnote}{0}

{\setlength\parindent{0em}%\noindent

M.~Bogomilov,  R.~Tsenov, G.~Vankova-Kirilova
\\{\it
   Department of Atomic Physics, St.~Kliment Ohridski University of Sofia, Sofia, Bulgaria
}\\

R.~Bertoni, M.~Bonesini, F.~Chignoli, R.~Mazza
\\{\it
Sezione INFN Milano Bicocca, Dipartimento di Fisica G.~Occhialini, Milano, Italy
}\\

V.~Palladino
\\{\it
Sezione INFN Napoli and Dipartimento di Fisica, Universit\`{a} Federico II, Complesso Universitario di Monte S.~Angelo, Napoli, Italy
}\\

A.~de Bari, G.~Cecchet
\\{\it 
Sezione INFN Pavia and Dipartimento di Fisica, Pavia, Italy
}\\

M.~Capponi, A.~Iaciofano, D.~Orestano, F.~Pastore\footnote{Deceased \label{dec}}, L.~Tortora
\\{\it
Sezione INFN Roma Tre e Dipartimento di Fisica, Roma, Italy
}\\

Y.~Kuno, H.~Sakamoto
\\{\it
Osaka University, Graduate School of Science, Department of Physics, Toyonaka, Osaka, Japan
}\\

S.~Ishimoto
\\{\it
High Energy Accelerator Research Organization (KEK), Institute of Particle and Nuclear Studies, Tsukuba, Ibaraki, Japan
}\\

F.~Filthaut\footnote{Also at Radboud University, Nijmegen, The Netherlands}
\\{\it
Nikhef, Amsterdam, The Netherlands
}\\

O.~M.~Hansen, S.~Ramberger, M.~Vretenar
\\{\it
CERN, Geneva, Switzerland
}\\

R.~Asfandiyarov, A.~Blondel, F.~Drielsma, Y.~Karadzhov 
\\{\it
DPNC, Section de Physique, Universit\'e de Gen\`eve, Geneva, Switzerland
}\\

G.~Charnley, N.~Collomb,  A.~Gallagher, A.~Grant, S.~Griffiths,  T.~Hartnett, B.~Martlew, A.~Moss, A.~Muir, I.~Mullacrane, A.~Oates, P.~Owens, G.~Stokes, P.~Warburton, C.~White
\\{\it
STFC Daresbury Laboratory, Daresbury, Cheshire, UK
}\\

D.~Adams, P.~Barclay, V.~Bayliss, T.~W.~Bradshaw, M.~Courthold, V.~Francis, L.~Fry, T.~Hayler, M.~Hills, A.~Lintern, C.~Macwaters, A.~Nichols, R.~Preece, S.~Ricciardi, C.~Rogers, T.~Stanley, J.~Tarrant, S.~Watson, A.~Wilson
\\{\it
STFC Rutherford Appleton Laboratory, Harwell Oxford, Didcot, UK
}\\
\\

R.~Bayes,  J.~C.~Nugent, F.~J.~P.~Soler\footnote{Corresponding author.}
\\{\it
School of Physics and Astronomy, Kelvin Building, The University of Glasgow, Glasgow, UK
}\\

P.~Cooke, R.~Gamet
\\{\it
Department of Physics, University of Liverpool, Liverpool, UK
}\\

A.~Alekou, M.~Apollonio, G.~Barber, D.~Colling, A.~Dobbs, P.~Dornan, C.~Hunt, J-B.~Lagrange, K.~Long, J.~Martyniak,  S.~Middleton, J.~Pasternak, E.~Santos, T.~Savidge, M.~A.~Uchida
\\{\it
Department of Physics, Blackett Laboratory, Imperial College London, London, UK
}\\

V.~J.~Blackmore\footnote{Now at Department of Physics, Blackett Laboratory, Imperial College London, London, UK},T.~Carlisle, J.~H.~Cobb, W.~Lau, M.~A.~Rayner, C.~D.~Tunnell
\\{\it
Department of Physics, University of Oxford, Denys Wilkinson Building, Oxford, UK
}\\

C.~N.~Booth, P.~Hodgson, J.~Langlands, R.~Nicholson, E.~Overton, M.~Robinson, P.~J.~Smith
\\{\it
Department of Physics and Astronomy, University of Sheffield, Sheffield, UK
}\\

A.~Dick, K.~Ronald, D.~Speirs, C.~G.~Whyte, A.~Young
\\{\it
Department of Physics, University of Strathclyde, Glasgow, UK
}\\

S.~Boyd,  P.~Franchini, J.~R.~Greis, C.~Pidcott, I.~Taylor
\\{\it
Department of Physics, University of Warwick, Coventry, UK
}\\

R.~Gardener, P.~Kyberd, M.~Littlefield, J.~J.~Nebrensky
\\{\it
Brunel University, Uxbridge, UK
}\\

A.~D.~Bross, T.~Fitzpatrick\footnotemark[1], M.~Leonova, A.~Moretti, D.~Neuffer, M.~Popovic, P.~Rubinov, R.~Rucinski
\\{\it
Fermilab, Batavia, IL, USA
}\\

T.~J.~Roberts
\\{\it
Muons, Inc., Batavia, IL, USA
}\\

D.~Bowring, A.~DeMello, S.~Gourlay, D.~Li, S.~Prestemon, S.~Virostek, M.~Zisman\footnotemark[1]
\\{\it
Lawrence Berkeley National Laboratory, Berkeley, CA, USA
}\\

M.~Drews, P.~Hanlet, G.~Kafka, D.~M.~Kaplan, D.~Rajaram, P.~Snopok, Y.~Torun, M.~Winter
\\{\it
Illinois Institute of Technology, Chicago, IL, USA
}\\

S.~Blot, Y.~K.~Kim
\\{\it
Enrico Fermi Institute, University of Chicago, Chicago, IL, USA
}\\

U.~Bravar
\\{\it
University of New Hampshire, Durham, NH, USA
}\\
\\

Y.~Onel
\\{\it
Department of Physics and Astronomy, University of Iowa, Iowa City, IA, USA
}\\

L.~M.~Cremaldi, T.~L.~Hart, T.~Luo, D.~A.~Sanders, D.~J.~Summers
\\{\it
University of Mississippi, Oxford, MS, USA
}\\

D.~Cline\footnotemark[1], X.~Yang
\\{\it
University of California, Los Angeles, CA, USA
}\\

L.~Coney, G.~G.~Hanson, C.~Heidt
\\{\it
University of California, Riverside, CA, USA
}\\

}

\end{document}